\documentclass[twocolumn,showpacs,superscriptaddress,amsmath,amssymb]{revtex4}
\usepackage{graphicx}% Include figure files
\usepackage{dcolumn}% Align table columns on decimal point
\usepackage{bm}% bold math
\def\bd{
\begin{document}} \def\ed{\end{document}}
\def\bmp{\begin{minipage}} \def\emp{\end{minipage}}
\def\bcc{\begin{center}} \def\ecc{\end{center}}     \def\npg{\newpage}
\def\beq{\begin{equation}} \def\eeq{\end{equation}} \def\hph{\hphantom}
\def\be{\begin{equation}} \def\ee{\end{equation}} \def\r#1{$^{[#1]}$}
\def\n{\noindent} \def\ni{\noindent} \def\pa{\parindent}
\def\hs{\hskip} \def\vs{\vskip} \def\hf{\hfill} \def\ej{\vfill\eject}
\def\cl{\centerline} \def\ob{\obeylines}  \def\ls{\leftskip}
\def\underbar#1{$\setbox0=\hbox{#1} \dp0=1.5pt \mathsurround=0pt
   \underline{\box0}$}   \def\ub{\underbar}    \def\ul{\underline}
\def\f{\left} \def\g{\right} \def\e{{\rm e}} \def\o{\over} \def\d{{\rm d}}
\def\vf{\varphi} \def\pl{\partial} \def\cov{{\rm cov}} \def\ch{{\rm ch}}
\def\la{\langle} \def\ra{\rangle} \def\EE{e$^+$e$^-$} \def\pt{p_{\rm t}}
\def\dt{\delta}
\def\bitz{\begin{itemize}} \def\eitz{\end{itemize}}
\def\btbl{\begin{tabular}} \def\etbl{\end{tabular}}
\def\btbb{\begin{tabbing}} \def\etbb{\end{tabbing}}
\def\beqar{\begin{eqnarray}} \def\eeqar{\end{eqnarray}}
\def\\{\hfill\break} \def\dit{\item{-}} \def\i{\item}
\def\bbb{\begin{thebibliography}{9} \itemsep=-1mm}
\def\ebb{\end{thebibliography}} \def\bb{\bibitem}
\def\bpic{\begin{picture}(260,240)} \def\epic{\end{picture}}
\def\akgt{\noindent{Acknowledgements}}
\def\fgn{\noindent{\bf\large\bf figure captions}}
%%%%%%%%%%%%%%%%  \begin{eqnarray*} \end{eqnarray}  %%%%%%%%%%%%%%%%
%%%%%%%%%%%%%%%%%%%%%%%%%%%%%%%%%%%%%%%%%%%%%%%%%%%%%%%%%%%%%%%%%%%%
\def\lan{\langle}
\def\ran{\rangle}
\def\p{\pi}
\def\ifmath#1{\relax\ifmmode #1\else $#1$\fi}%
\def\rc{\ifmath{{\mathrm{c}}}}
\def\cut{\ifmath{{\mathrm{cut}}}}
\def\rF{\ifmath{{\mathrm{F}}}}
\def\rK{\ifmath{{\mathrm{K}}}}
\def\rp{\ifmath{{\mathrm{p}}}}
\def\rt{\ifmath{{\mathrm{t}}}}
\def\LAB{\ifmath{{\mathrm{LAB}}}}
\def\cut{\ifmath{{\mathrm{cut}}}}
\def\beq{\begin{equation}}
\def\eeq{\end{equation}}

\newcommand{\cinst}[2]{$^{\mathrm{#1}}$~#2\par}
\newcommand{\crefi}[1]{$^{\mathrm{#1}}$}
\newcommand{\crefii}[2]{$^{\mathrm{#1,#2}}$}
\newcommand{\crefiii}[3]{$^{\mathrm{#1,#2,#3}}$}
\newcommand{\HRule}{\rule{0.5\linewidth}{0.5mm}}

\bd

\title{Boost Invariance and Multiplicity Dependence  \\of the Charge Balance Function
 in $\pi^{+}\rp$ and $\rK^{+}\rp$ Collisions at $\sqrt s= 22$ GeV/$c$}

\author{M.R.~Atayan}
\affiliation{ Institute of Physics, AM-375036 Yerevan, Armenia}

\author{Bai Yuting}
\affiliation{ Institute of Particle Physics, Hua-Zhong Normal
University, Wuhan 430070, China}

\author{E.A.~De Wolf}
\affiliation{ Department of Physics, University of Antwerp, B-2610
Wilrijk, Belgium}

\author{A.M.F.~Endler}
\affiliation{ Centro Brasileiro de Pesquisas Fisicas, BR-22290 Rio
de Janeiro, Brazil}

\author{Fu Jinghua}
\affiliation{ Institute of Particle Physics, Hua-Zhong Normal
University, Wuhan 430070, China}

\author{H.~Gulkanyan}
\affiliation{ Institute of Physics, AM-375036 Yerevan, Armenia}

\author{R.~Hakobyan}
\affiliation{ Institute of Physics, AM-375036 Yerevan, Armenia}

\author{W.~Kittel}
\affiliation{ Radboud University Nijmegen/NIKHEF, NL-6525~ED
Nijmegen, The Netherlands}

\author{Liu Lianshou}
\affiliation{ Institute of Particle Physics, Hua-Zhong Normal
University, Wuhan 430070, China}

\author{Li Zhiming}
\affiliation{ Institute of Particle Physics, Hua-Zhong Normal
University, Wuhan 430070, China}

\author{Li Na}
\affiliation{ Institute of Particle Physics, Hua-Zhong Normal
University, Wuhan 430070, China}

\author{Z.V.~Metreveli}
\affiliation{ Institute for High Energy Physics of Tbilisi State
University, GE-380086 Tbilisi, Georgia; now at Northwestern Univ.,
Evanston, U.S.A.}

\author{L.N.~Smirnova}
\affiliation{ Scobeltsyn Institute of Nuclear Physics, Lomonosow
Moscow State University, RU-119899 Moscow, Russia}

\author{L.A.~Tikhonova}
\affiliation{ Scobeltsyn Institute of Nuclear Physics, Lomonosow
Moscow State University, RU-119899 Moscow, Russia}

\author{A.G.~Tomaradze}
\affiliation{ Institute for High Energy Physics of Tbilisi State
University, GE-380086 Tbilisi, Georgia; now at Northwestern Univ.,
Evanston, U.S.A.}

\author{Wu Yuanfang}
\affiliation{ Institute of Particle Physics, Hua-Zhong Normal
University, Wuhan 430070, China}

\author{S.A.~Zotkin}
\affiliation{ Scobeltsyn Institute of Nuclear Physics, Lomonosow
Moscow State University, RU-119899 Moscow, Russia}
\affiliation{Now at DESY, Hamburg, Germany}

\collaboration{EHS/NA22 Collaboration}

\begin{abstract}

Boost invariance and multiplicity dependence of the charge balance
function are studied in $\pi^{+}\rp$ and $\rK^{+}\rp$ collisions
at 250 GeV/$c$ incident beam momentum. Charge balance, as well as
charge fluctuations, are found to be boost invariant over the
whole rapidity region, but both depend on the size of the rapidity
window. It is also found that the balance function becomes
narrower with increasing multiplicity, consistent with the
narrowing of the balance function when centrality and/or system
size increase, as observed in current relativistic heavy ion
experiments.

\end{abstract}


\maketitle Charge balance and charge flow are measures of rapidity
correlations between oppositely charged particles and have been
used to study hadronization in hadron-hadron~\cite{oldbf} as well
as in lepton-hadron~\cite{lh} and e$^+$e$^-$~\cite{e+e-}
collisions. In the form of a charge balance function
(BF)~\cite{bf1}, they have recently gained new interest in the
field of relativistic heavy ion collisions. A narrowing of the
balance function is suggested as a signature of delayed
hadronization, or of the formation of a Quark-Gluon Plasma (QGP)
during the early stage of a collision. The integral of the balance
function is related to the event-by-event charge
fluctuations~\cite{bf2}, which are expected to be suppressed in a
QGP~\cite{charges}.

So far, two heavy ion experiments~\cite{star, na49} have measured
the balance function at various centralities and for different
colliding nuclei. A narrowing of the balance function is indeed
observed with increasing centrality of the collision and with
increasing size of the colliding nuclei.
Coalescence~\cite{bialas}, thermal~\cite{thermal} or blast
wave~\cite{cheng} models with transverse flow and local clusters
and/or resonances can explain these phenomena. The measured charge
fluctuations, on the other hand, are consistent with those
expected for a hadronic gas~\cite{oview,cstar, cphenix}.

Before drawing any conclusions from the observed narrowing of the
balance function, it is necessary to know how the BF behaves in
hadron-hadron collisions, where no QGP is expected, and how the
limited detector acceptance influences its
width~\cite{bialas,bf1,bf2}. Since current heavy ion
experiments~\cite{star, cphenix, na49} cover only a limited
rapidity region, the measured BF's do not correspond to that for
the full rapidity region. Whether the results from different heavy
ion experiments are comparable or not depends on the influence of
the acceptance.

A number of theoretical
discussions~\cite{acep1,acep2,charges,ccharge,bf2} have been
devoted to the influence of acceptance, or on how to find robust
measures for the charge correlations and fluctuations. In
particular, in ref.~\cite{bf2}, based on the assumption of
longitudinal boost invariance (rapidity independence), Jeon and
Pratt proposed a relation between the balance function in a
rapidity window $B(\dt y|Y_{\rm w})$ and in the full rapidity
range $B(\dt y|Y=\infty)$~\cite{bf2},
\beqar  %%(1)
 B(\dt y|Y_{\rm w})=B(\dt y|\infty)\left( 1-\frac{\dt y}{Y_{\rm w}}\right),
\eeqar \noindent where $Y_{\rm w}$ is the size of the rapidity
window and $B(\dt y|Y_{\rm w})$ can be measured by
\begin{eqnarray}
 B(\dt y|Y_{\rm w}) &=& \frac{1}{2}\left[
 \frac{\la n_{+-}(\dt y)\ra-\la n_{++}(\dt y)\ra}{\la
 n_+\ra}\right.    %%%\right]\end{eqnarray}
 \nonumber \\
 & &\quad \left.+\frac{\la n_{-+}(\dt y)\ra-\la n_{--}(\dt y)\ra}{\la
n_-\ra}\right].\end{eqnarray} Here, $n_{+-}(\dt y)$, $n_{++}(\dt
y)$ and $n_{--}(\dt y)$ are the numbers of pairs of opposite- and
like-charged particles satisfying the criteria that they fall into
the rapidity window $Y_{\rm w}$ and that their relative rapidity
equals $\dt y$; $n_+$ and $n_-$ are the numbers of positively and
negatively charged particles, respectively, in the interval
$Y_{\rm w}$.

Conventionally, boost invariance refers to particle density being
independent of rapidity, as originally assumed in~\cite{Fey} and
applied in a simple solvable hydrodynamic model~\cite{hydro, bj}.
While this may be correct in a very restricted region at
mid-rapidity for the {\it rapidity density} itself~\cite{NA22,UA5,
CDF, AA, BiJa}, boost invariance of the {\it balance function}
only requires that the {\it charge correlation} between final
state particles be the same in any
longitudinally-Lorentz-transformed frame. Whether the BF is boost
invariant over the {\it whole} rapidity region or only in the
central region, cannot be simply deduced from the corresponding
shape of the rapidity density distribution. This important issue
has not yet been investigated in either its theoretical or
experimental aspects.

In this letter, boost invariance and multiplicity dependence of
the charge balance function is studied on $\pi^{+}\rp$ and
$\rK^{+}\rp$ data at 250 GeV/$c$ ($\sqrt s=$22 GeV) of the NA22
experiment. This experiment was equipped with a rapid cycling
bubble chamber as an active vertex detector, had excellent
momentum resolution and $4\pi$ acceptance. The latter feature
allows, for the first time, to study the properties of the balance
function in full phase space.

Since no statistically significant differences are seen between
the results for $\pi^{+}$ and $\rK^{+}$ induced reactions, the two
data samples are combined for the purpose of this analysis. A
total of 44~524 non-single-diffractive events is obtained after
all necessary selections, as described in detail
in~\cite{na22data}. In particular, possible contamination from
secondary interactions is suppressed by a visual scan and the
requirement that overall charge balance be satisfied within the
whole event; $\gamma$ conversions near the primary vertex are
removed by electron identification.

In Fig.~1, the balance function is shown for five cms rapidity
windows of width $Y_{\rm w}=3$, located at different positions,
$[-3,0]$ (open stars), $[-2, +1]$ (open crosses), $[-1.5, 1.5]$
(open circles), $[-1, 2]$ (open diamonds), and $[0, 3]$ (open
triangles). In this and the following figures, errors are smaller
than the size of the symbols. The five functions coincide within
the experimental errors, except that a few points in $[-3, 0]$ are
somewhat lower than the others. This is caused by very low
multiplicities in the rapidity region $[-3, -2]$, where
unidentified protons contribute and where the rapidity
distribution is not completely symmetric to the region $[2, 3]$.
The figure demonstrates that, despite a strong rapidity dependence
of the particle density given in Fig.~2, the balance function is
largely independent of the position of the rapidity window, i.e.,
the charge correlation is essentially the same in any
longitudinally-Lorentz-transformed frame.

\begin{figure}
\includegraphics[width=2.2in]{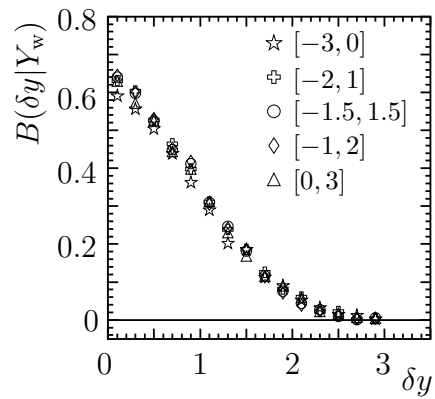}% Here is how to import EPS art
\caption{\label{Fig. 1} The balance function for five different
positions of a rapidity window of size $Y_{\rm w}=3$: $[-3, 0]$
(open stars), $[-2, 1]$ (open crosses), $[-1.5, 1.5]$ (open
circles), $[-1, 2]$ (open diamonds) and $[0, 3]$ (open
triangles).}
\end{figure}

\begin{figure}
\includegraphics[width=2.2in]{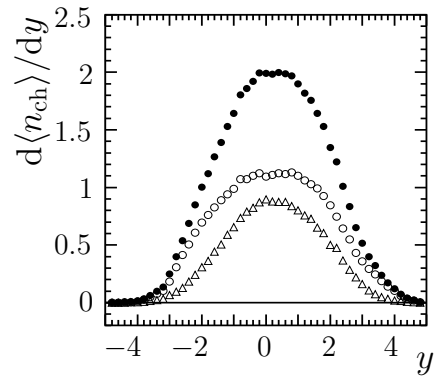}% Here is how to import EPS art
\caption{\label{Fig. 2} Center-of-mass rapidity distribution of
positively (open circles), negatively (open triangle), and all
charged (solid circles) particles.}
\end{figure}

Since boost invariance of the BF is found to be valid over the
whole rapidity region, it is now interesting to verify if the BF
in a limited rapidity window can be deduced from that in the full
rapidity region by Eq.~(1), and vice versa. In Fig.~3, the balance
function, $B(\dt y|Y_{\rm w})$ (solid points), for four rapidity
windows (central in Fig.~3a, non-central in Fig.~3b),  is compared
to $B(\dt y|\infty)(1-\frac{\dt y}{Y_{\rm w}})$ (open points)
obtained for the corresponding window from the BF in the full
region. The data confirm that the relation Eq.~(1) is indeed
approximately satisfied, independently of size or position of the
window. This result is especially useful for experiments with
limited acceptance, in particular for the current heavy ion
experiments.

Fig.~3 further illustrates that the BF becomes narrower with
decreasing $Y_{\rm w}$, in agreement with Eq.~(1).

\begin{figure}
\includegraphics[width=3.2in]{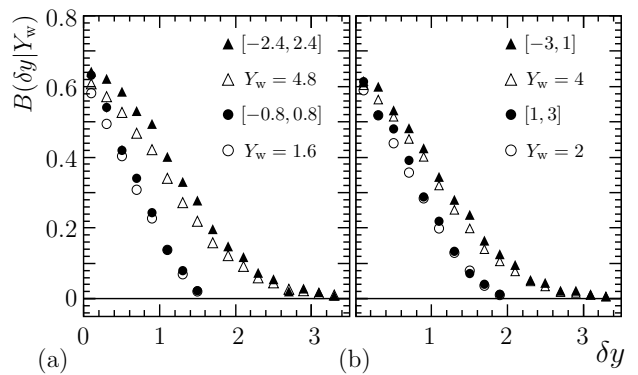}% Here is how to import EPS art
\caption{\label{Fig. 3} The balance functions $B(\dt y|Y_{\rm w})$
(solid symbols) (a) for two central rapidity windows,  $[-2.4,
2.4]$ (triangles) and $[-0.8, 0.8]$ (circles) and (b) two
asymmetric rapidity windows $[-3, 1]$ (triangles), and $[1, 3]$
(circles), compared with corresponding $B(\dt y|\infty)\cdot
(1-\frac{\dt y}{Y_{\rm w}})$ (open symbols).}
\end{figure}

Since the charge fluctuation $D(Q)$~\cite{bf2} is approximately
related to the BF by
\beqar  %%(1)
 \frac{D(Q)}{4}=1-\int_0^{Y_{\rm w}}B(\dt y|Y_{\rm w})\mathrm{d}\delta
 y+\mathcal{O}\left(\frac{\langle Q\rangle}{\langle
 n_{\mathrm{ch}}\rangle}\right),
\eeqar \noindent where $Q=n_+-n_-$ and $n_{\mathrm{ch}}=n_++n_-$,
it is interesting to see how the charge fluctuation changes with
position and size of the rapidity window. For this purpose,
$D(Q)/4$ is presented in Fig.~4 for different positions and sizes
of a rapidity window, $Y_{\rm w}=1.0$ (circles), 2.0 (triangles),
and 3.0 (stars). The results confirm that for a given window size
its value is independent of the position of that
window~\cite{na22c}, in agreement with the boost invariance of the
balance function.  The data also show that $D(Q)$ is sensitive to
the size of the observed window. So it is necessary to give the
exact size of the rapidity region when the fluctuation is treated
quantitatively~\cite{charges}.

As has been demonstrated in~\cite{na22c}, $D(Q)$ also depends on
the acceptance in transverse momentum and azimuthal angle. In
Fig.~4, $D(Q)/4$ is also presented under the same transverse momentum
and azimuthal
angle cuts as used in STAR ($p_{\rm t}>0.1$ GeV/c) and PHENIX ($p_{\rm
t}>0.2$ GeV/c and $\Delta\phi=\pi/2$) with a rapidity
window of size $Y_{\rm w}=1.0$, by open circles and open
squares, respectively. The transverse momentum cut used by STAR has little
influence on the result, while the combined transverse momentum and
azimuthal cut used by PHENIX destroys the boost invariance
of $D(Q)$. These results show that a limited acceptance
in transverse momentum and azimuthal angle can destroy the
boost invariance of charge fluctuations. Furthermore, it has been
verified that the effect is the larger
the larger the percentage of particles lost.

\begin{figure}
\includegraphics[width=2.2in]{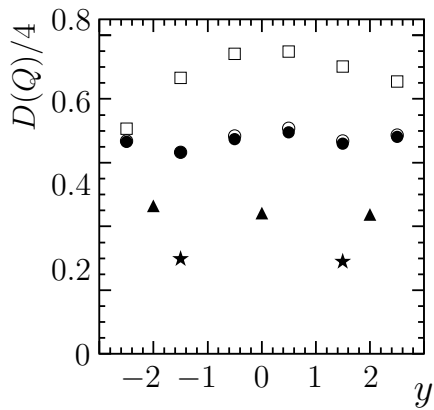}% Here is how to import EPS art
\caption{\label{Fig. 4} $D(Q)/4$ versus the position of a rapidity
window of size $Y_{\rm w}=1$ (circles), 2 (triangles), and 3
(stars). Open circles and open squares are $D(Q)/4$ under the same
transverse momentum and azimuthal angle cuts as STAR ($p_{\rm
t}>0.1$ GeV/c) and PHENIX ($p_{\rm t}>0.2$ GeV/c and
$\Delta\phi=\pi/2$) with a rapidity window of size $Y_{\rm
w}=1.0$. }
\end{figure}

In Fig.~5, the full-rapidity BF, $B(\dt y|\infty)$, is presented
for all charged particles (full circles) and for the three
multiplicity intervals, $0 < n_{\mathrm{ch}} < 6$ (open circles),
$6 \le n_{\mathrm{ch}} \le 8$ (open triangles), and
$n_{\mathrm{ch}}> 8$ (open stars). The width of the balance
function, defined as \beqar \langle \delta y\rangle=\frac{\sum_i
B( \delta y_i|\infty)\delta y_i}{\sum_i B( \delta
y_i|\infty)},\eeqar for the corresponding multiplicity intervals
and for all charged particles is listed in Table I. The width
decreases with increasing multiplicity. This is, at least
qualitatively, consistent with the narrowing of the balance
function with increasing centrality observed in current heavy ion
experiments~\cite{star, na49}. So, before a narrowing of the BF
with increasing centrality and increasing mass number of the
colliding nuclei can be interpreted as due to the formation of a
QGP, the multiplicity effect observed here should be properly
accounted for.

\begin{figure}
\includegraphics[width=2.2in]{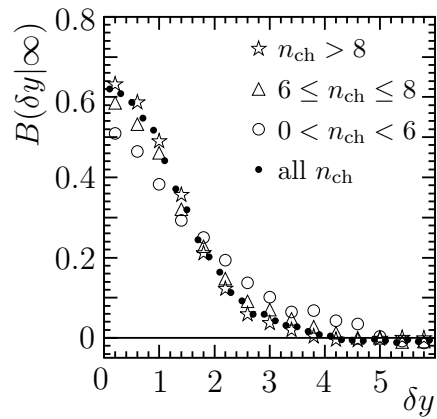}% Here is how to import EPS art
\caption{\label{Fig. 5} The balance function for all charged
particles (full circles) and for three multiplicity intervals, $0
< n_{\mathrm{ch}} < 6$ (open circles), $6 \le n_{\mathrm{ch}} \le
8$ (open triangles) and $n_{\mathrm{ch}}>8$ (open stars).}
\end{figure}

\begin{table}
\caption{\label{Table 1.} The width of the balance function in
three \\  multiplicity intervals and for all charged particles}
\begin{tabular}{ccc} \hline\small
  Multiplicity & &$\langle \delta y\rangle$ \\ \hline $n_{\rm ch} >8$
&& 0.957 $\pm$ 0.011  \\
$6\leq n_{\rm ch} \leq 8$& & 1.096 $\pm$ 0.014 \\
$0< n_{\rm ch} < 6$ && 1.359 $\pm$ 0.026 \\
all $n_{\rm ch}$ & &0.991 $\pm$ 0.008 \\
\hline
\end{tabular}
\end{table}

The results of this paper can be summarized as follows:

1. In contrast to the strong dependence of the particle density on
rapidity, the BF is invariant under a longitudinal boost over the
whole rapidity region. This property allows to determine the BF in
full rapidity, $B(\delta y|\infty)$, from a measurement with
limited rapidity acceptance.

2. The balance function becomes narrower with decreasing size of
the window. Therefore, only the full-rapidity BF can be used in
comparing data from different experiments.

3. The balance function becomes narrower with increasing
multiplicity, an effect also observed in heavy ion interactions
when the centrality of the collision increases.

4. The charge fluctuations are boost invariant but depend on the
size of the rapidity window.

\vs 5mm

This work is part of the research program of the "Stichting voor
Fundamenteel Onderzoek der Materie (FOM)", which is financially
supported by the "Nederlandse Organisatie voor Wetenschappelijk
Onderzoek (NWO)". We further thank NWO for support of this project
within the program for subsistence to the former Soviet Union
(07-13-038). The Yerevan group activity is financially supported,
in the framework of the theme No. 0248, by the Government of the
Republic of Armenia. This work is also supported in part by the
National Natural Science Foundation of China with project
No.10375025 and No.10475030 and the Ministry of Education of China
with project No.Jiaojisi(2004)295 and by the Royal Dutch Academy
of Sciences under the Project numbers 01CDP017, 02CDP011 and
02CDP032 and by the U.S. Department of Energy.

\ed